\documentclass{iopart}
\usepackage{epsfig}

\begin{document}
\jl{4}

\def\bq{\begin{eqnarray}}
\def\eq{\end{eqnarray}}
\def\l{\langle}
\def\r{\rangle}

\title{QCD corrections to $e^+ e^- \rightarrow 4 \; \mbox{jets}$}

\author{S Weinzierl\footnote[1]{Talk given at the UK Phenomenology
    Workshop on Collider Physics, Durham, 19-24 September 1999}}
\address{NIKHEF, P.O. Box 41882, 1009 - DB Amsterdam, The Netherlands}


\begin{abstract}
We report on the next-to-leading order QCD calculation for
$e^+ e^- \rightarrow 4 \; \mbox{jets}$.
We explain some modern techniques which have been used to calculate the one-loop
amplitudes efficiently.
We further report on the general purpose numerical program ``Mercutio'', which can
be used to calculate any infrared safe four-jet quantity in electron-positron
annihilation at next-to-leading order.
\end{abstract}

\section{Motivation: LEP physics}

QCD four-jet production in $e^+e^-$ annihilation can be measured at LEP and can be studied in its own
right. First of all, $e^+ e^- \rightarrow \mbox{4 jets}$ is the lowest order process which contains
the non-abelian three-gluon-vertex at tree level and thus allows for an measurement of the
colour factors $C_F$, $C_A$ and $T_R$ of QCD. This in turn may be used to put exclusion limits on 
light gluinos.
Furthermore the QCD process is a background to W-pair production, when both W's decay
hadronically and to certain channels of the search of the Higgs boson like
$e^+ e^- \rightarrow Z^{\ast} \rightarrow Z H \rightarrow \mbox{4 jets}$.
The one-loop matrix elements required for an NLO study of four-jet production
are also an essential input for an NNLO calculation of three-jet production.
The latter one would be needed to reduce theoretical uncertainties in the extraction
of the strong coupling at the Z-pole.
 
In general leading-order calculations in QCD give a rough description of the process under consideration, but
they suffer from large uncertainties. The arbitrary choice of the renormalization scale gives rise to an ambiguity, which is 
reduced only in an next-to-leading order calculation. Furthermore the internal structure of a jet and
the sensitivity to the merging procedure of the jet algorithm are modelled only in an NLO
analysis. Both uncertainties are related to the appearance of logarithms, ultraviolet in
nature in the first case, infrared in the latter, which are calculated explicitly only in an
NLO calculation.

A NLO calculation proceeds in two steps: First, one needs the relevant amplitudes, in our case
$e^+ e^- \rightarrow 4 \;\mbox{partons}$ and
$e^+ e^- \rightarrow 5 \;\mbox{partons}$ at tree level and $e^+ e^- \rightarrow 4 \;\mbox{partons}$
at one loop. 
Among these, the one-loop amplitudes are the most complicated ones and we will comment
on their calculation in the next section.
The second step requires setting up a numerical Monte Carlo program which has to deal with
infrared divergences. We will focus on this point in the third section.
In the last section we will give some numerical results.

\section{One-loop amplitudes}

We used a variety of modern techniques 
in order to calculate the one-loop amplitudes efficiently. 
These include colour decomposition, where
amplitudes are decomposed into simpler gauge-invariant partial amplitudes with 
definite colour structure, and the spinor helicity method, which consists in expressing all
Lorentz four-vectors and Dirac spinors in terms of massless two-component Weyl-spinors.
Their use divides the
task into smaller, more manageable pieces. Also a decomposition inspired by
supersymmetry proved to be useful, where the particles running around the loop are reexpressed
in terms of supermultiplets.
In a second step the cut technique and factorization in collinear limits
are used to constrain the analytic form of the partial amplitudes.

As an example we explain in more detail the cut technique \cite{Bern:1995cg}, which is based on unitarity.
To obtain the coefficients of the basic box, triangle or bubble integrals one considers the cuts
in all possible channels. Each phase-space integral is rewritten with the help of the
Cutkosky rules as the imaginary part of a loop amplitude. The power of this method lies
within the fact, that on each side of the cut one has a full tree amplitude and not just a single
Feynman diagram. This method allows one to reconstruct the one-loop amplitude up to terms
without an imaginary part. The remaining terms were obtained by
examining the collinear limits.

For the reduction of tensor pentagon integrals we used a new reduction algorithm \cite{Weinzierl:1998we},
based on the Schouten identity and Weyl spinors, which does not introduce artifical Gram
determinants in the denominator.

The one-loop amplitudes for the first subprocess 
$e^+e^- \rightarrow q \bar{q} Q \bar{Q}$ were calculated in refs.
\cite{Bern:1997ka,Glover:1997eh} and the amplitudes for the second subprocess $e^+e^-$ $\rightarrow q \bar{q} g g$
in refs. \cite{Bern:1997sc,Campbell:1997tv}. The calculations of the
two groups agree with each other.

\section{Numerical implementation: Mercutio}

The second major part of a general purpose NLO program for four jets is coding
the one-loop amplitudes and the five parton tree-level amplitudes
in a numerical Monte Carlo program. 
At leading order the task is relatively simple: One parton corresponds to one jet. At
NLO however, a jet can be modeled by two partons. At NLO the cross section receives
contributions from the virtual corrections and the real emission part.
Only the sum of them is infrared finite, whereas when taken separately, 
each part gives a divergent contribution.
Several methods to handle this problem exist, such as the phase-space slicing method
\cite{Giele:1992vf}, the subtraction method \cite{Frixione:1996ms} and the dipole formalism \cite{Catani:1997vz}.
We have chosen the dipole formalism.
Within the dipole formalism one subtracts and adds again a suitably chosen term:
\bq
\sigma^{NLO} & = & \int\limits_{n+1} \left( d\sigma^R - d\sigma^A \right)
    + \int\limits_n \left( d\sigma^V + \int\limits_1 d\sigma^A \right)
\eq
The approximation term $d\sigma^A$ has to fullfill the following two requirements:
First, $d\sigma^A$ must be a proper approximation to $d\sigma^R$, with the same
pointlike singular behaviour in $D$ dimensions as $d\sigma^R$.
Secondly, $d\sigma^A$ must be analytically integrable in $D$ dimensions
over the one-parton subspace leading to the soft and collinear divergences.

Let me now turn to the details of the Monte Carlo integration.
The heart of any Monte Carlo integration is the random number generator.
Among other things, it should have a long period and should not introduce
artifical correlations. As the default random number generator we use
\bq
s_i & = & \left( s_{i-24} + s_{i-55} \right) \; \mbox{mod} \; 2^{32}.
\eq
It was proposed by Mitchell and Moore and has a period of $2^f(2^{55}-1)$,
where $0\le f \le 32$.
Massless fourmomenta are generated with the help of the RAMBO-algorithm
\cite{Kleiss:1986gy}. This algorithm generates events with a uniform
weight.
Adaptive importance sampling is implemented using the VEGAS-algorithm \cite{Lepage:1978sw}.
A naive implementation of the dipole
formalism will give large statistical errors when performing a Monte
Carlo integration over the real corrections with dipole
factors subtracted. In order to improve the efficiency of the Monte Carlo
integration we remap the phase space to make the integrand 
more flat.
A simplified model for the term $d\sigma^R - d\sigma^A$ would be
\bq
\label{model}
F & = & \int\limits_0^1 dx \left( \frac{f(x)}{x} - \frac{g(x)}{x} \right)
\eq
where $f(0) = g(0)$ is assumed.
$f(x)/x$ corresponds to the
original real emission part with a soft or collinear singularity at
$x=0$, $g(x)/x$ corresponds to the subtraction term of the dipole
formalism.
\def\ymin{y_{\rm min}}
Eq. (\ref{model}) can be rewritten as
\bq
F & = & \int\limits_0^{\ymin} dx \frac{f(x)-g(x)}{x} 
 + \int\limits_{\ln \ymin}^0 dy \left( f(e^y) - g(e^y) \right)\,,
\eq
where $\ymin$ is an artificial parameter separating a numerically
dangerous region from a stable region.  Using the Taylor expansion for
$f(x)-g(x)$, one sees that the first term gives a contribution of
order $O(\ymin)$.  In the second term the $1/x$ behaviour has been
absorbed into the integral measure by a change of variables $y = \ln x$,
and the integrand tends to be more flat. 
It should be noted that there is no approximation involved.

\section{Numerical results}

Numerical programs for $e^+ e^- \rightarrow 4 \; \mbox{jets}$
have been provided by four groups: MENLO PARC \cite{Dixon:1997th},
DEBRECEN \cite{Nagy:1998bb}, EERAD2 \cite{Campbell:1998nn}
and MERCUTIO \cite{Weinzierl:1999yf}.
Various cross-checks have been performed among these programs and they agree
within statistical errors.
Here we report on the numerical program ``Mercutio'', which was written
in C++.
The four-jet fraction is defined as
\begin{eqnarray}
R_4 & = & \frac{\sigma_{4-jet}}{\sigma_{tot}}.
\end{eqnarray}
The values obtained for the four-jet fraction for the DURHAM algorithm with $y_{cut}=0.01$ for
various energies are given in table \ref{4jetfrac}.
\begin{table}
\begin{center}
\begin{tabular}{|c|c|c|}
\hline
$\sqrt{Q^2}$ & $R_4^{LO}$ & $R_4^{NLO}$ \\
\hline
$m_Z$   & $(2.98 \pm 0.01) \cdot 10^{-2} $ & $(4.72 \pm 0.01) \cdot 10^{-2} $ \\
135 GeV & $(2.65 \pm 0.01) \cdot 10^{-2} $ & $(4.12 \pm 0.01) \cdot 10^{-2} $ \\
161 GeV & $(2.53 \pm 0.01) \cdot 10^{-2} $ & $(3.89 \pm 0.01) \cdot 10^{-2} $ \\
172 GeV & $(2.48 \pm 0.01) \cdot 10^{-2} $ & $(3.81 \pm 0.01) \cdot 10^{-2} $ \\
183 GeV & $(2.44 \pm 0.01) \cdot 10^{-2} $ & $(3.73 \pm 0.01) \cdot 10^{-2} $ \\
189 GeV & $(2.42 \pm 0.01) \cdot 10^{-2} $ & $(3.69 \pm 0.01) \cdot 10^{-2} $ \\
\hline
\end{tabular}
\end{center}
\caption{\label{4jetfrac}
The four-jet fraction at LO and NLO for the DURHAM algorithm with $y_{cut}=0.01$ and
various energies.}
\end{table}
The decrease with energy is mainly due to the running of the strong coupling.

With the numerical program for $e^+ e^- \rightarrow \mbox{4 jets}$ one may also 
study the internal
structure of three-jets events. One example is the jet broadening variable
defined as 
\begin{eqnarray}
B_{jet} & = & \frac{1}{n_{jets}} \sum\limits_{jets} \frac{\sum\limits_{a} |p_a^\perp |}{\sum\limits_a |\vec{p}_a|}
\end{eqnarray}
Here $p^\perp_a$ is the momentum of particle $a$ transverse to the jet axis of jet $J$, and the sum over $a$ extends over
all particles in the jet $J$. 
The jet broadening variable is calculated for three-jet events defined by the DURHAM algorithm and $y_{cut} = 0.1$. This choice is motivated by a recent analysis of the Aleph collaboration \cite{Barate:1998cp}.
\begin{figure}
\centerline{
\epsfig{file=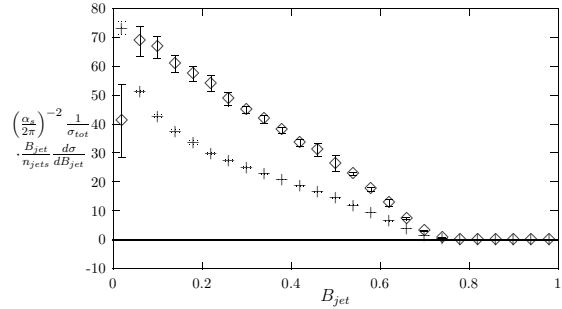,
bbllx=75pt, bblly=320pt, bburx=485pt, bbury=540pt,
width=8cm
}
}
\caption{\label{figa} The $B_{jet}$ distribution at NLO (diamonds) and LO (crosses).}
\end{figure}
Figure \ref{figa} shows the distribution of the jet broadening variable.

\section*{References}


\end{document}